\documentclass[12pt]{iopart}

\usepackage{graphicx}
\usepackage{hyperref}
\usepackage{listings} 
\begin{document}

\title{Particle-hole symmetry numbers for nuclei}
\author{Ian Bentley}
\address{Dept. of Chemistry and Physics, Saint Mary's College, IN 46556, Notre Dame, USA} 

{\bf Abstract:} 
Two new numbers, $\nu$ and $\zeta$, inspired by particle-hole symmetry are introduced. These numbers have extreme values at a closed shell and vanish mid-shell. A combination of even powers of these numbers has been used to model
experimentally measured quantities such as $R_{4/2}=E(4_1^+)/E(2_1^+)$   
and the \textquotedblleft microscopic" contribution to binding energies. A binding
energy fit consisting of a total of six fit coefficients, 
including one new shell term, reproduces the experimental binding energies of 2353
nuclei with an r.m.s. standard deviation of 1.55 MeV. 
The difference between the experimental and fit values of observables,
specifically the $R_{4/2}$, provides an indication of where shell closure features are less pronounced
 and where sub-shells closures occur. \\



\maketitle

\section{Introduction}
Electron-hole symmetry has been used to model half-filled semiconductor systems, e.g.
\cite{PHPRLmodel}, \cite{PHPRBmodel}. 
This symmetry is often used as a simplification by treating the lack of an 
electron below the Fermi surface and particle above the Fermi surface similarly \cite{PHnanoNature}.
In this paper, particle-hole symmetry is invoked to generate valence numbers for protons and neutrons.
These numbers have various possible uses for the modeling features related to shell structure.

Shell structure in nuclei is analogous to atomic shell structure that provides the 
basis for the periodic table \cite{Talmi}. 
Nuclear shell structure was first
seen in
the large number of stable nuclei at certain  \textquotedblleft magic$\textquotedblright$ numbers \cite{M48}, \cite{HJS49}, that are now known
to be $2,8,20,28,50,82,(126)$ for protons and neutrons.  Experimentally
measured energies of excited states, transition rates, 
have been known to indicate shell structure for more than 60 years, see e.g. \cite{FH49},\cite{Pryce54}.

The Interacting Boson Model (IBM) \cite{AI75}, 
which treats pairs of valence protons and neutrons as bosons, sets a excellent foundation for modeling shell structure. 
There are several IBM-based formalisms, that are correlated to shell structure. 
Strengths of using an IBM-based formalism are that they contain operators which utilize various 
symmetries, see e.g. \cite{IBMReview} or \cite{IBMReview2}, and they provide the ability to 
classify nuclear properties like the low lying spectra \cite{CastenNature}. A new formalism motivated by particle-hole symmetry
is introduced in the following section that is seemingly comparable to two bodies of work that exist within the framework set by the IBM.

The promiscuity factor ($P$) is one of the best examples of an IBM based quantity
used to model experimental observation.
It is defined as $P=\frac{N_p N_n}{N_p+N_n}$,
where $N_p$ and $N_n$ are counted as the numbers of particles or holes,
 depending on if the shell is less
than or more than half full, for
protons and neutrons respectively \cite{CastenP85}. 
In some sense $P$ serves as a measure of the strength of proton-neutron
interactions per valence nucleon. 

The ratio of excitation energies $R_{4/2}=E(4_1^+)/E(2_1^+)$ in even-even
nuclei is seen to vary depending on the shape of the nucleus. 
Small values of $R_{4/2}$ occur for
closed shell, spherical nuclei, and the larger values occur for well
deformed nuclei in mid-shell regions.
Small values of the $P$ factor have been shown to correspond to low values of $R_{4/2}$, 
and $R_{4/2}$ increases at  a point near $P\approx3$ or $P\approx4$ \cite{CastenP85}. The understanding is that this transition corresponds to the onset of deformation.

Shell structure can also be seen in experimental binding energies where nuclei with a closed shell 
are often more tightly bound together than are neighboring nuclei with no closures primarily as a result of deformation effects. 
Furthermore, nuclei with closed shells of both protons and neutrons are even more tightly bound
than those with just one shell closure as a result of enhanced pairing correlations from both sets of nucleons. The $P$ factor is insufficient in describing this because it is unable to distinguish between singly and doubly magic nuclei.

Dieperink, Van Isacker and others, have been able to model shell features in binding energies by introducing two IBM-based terms 
proportional to $(N_p+N_n)$ and $(N_p+N_n)^2$ \cite{VanI07}. These result in a root mean squared standard deviation ($\sigma$) of 1.4 MeV using just eight adjustable parameters. After modifying 
the location of three shell closures a fit with $\sigma=$1.2 MeV was achieved \cite{VanIsackerGlobal}.
This manuscript contains a further reduction of the number of adjustable coefficients while providing a semi-empirical fit with a comparable degree of accuracy. 

In Sections \ref{phsym}-\ref{AltRep}, the motivation for the numbers is introduced, and definitions of the numbers are provided.  Sections \ref{BE2test}-\ref{Sec.fittype}, contain initial tests and applications of the formalism and Section \ref{outlook} contains a discussion of the results and an outlook.

\section{Particle-hole symmetry} \label{phsym}

Particle-hole conjugation has been used by Bell \cite{Bell59} in the context of the shell model and extended further by M\"{u}ller-Arnke \cite{MA73}.
Davis et al. have employed a measure of the deviation from particle-hole conjugation symmetry of the form:
\begin{equation} \label{rho}
 \rho=\frac{| E_{pp}-E_{hh} |}{E_{pp}+E_{hh}},
\end{equation}
to test the validity of F-spin multiplets \cite{Davis91}. Here $E_{pp}$ corresponds to the energy of an excited state in a nucleus with valence protons and neutrons, and $E_{hh}$
corresponds to the energy of a state in a nucleus with proton and neutron holes. The finding was that pairs of $E(2_1^+)$ and $E(4_1^+)$ states in an F-spin multiplet deviate from
one another on the order of $10-20\%$ \cite{Davis91}. 

\begin{figure}
\begin{center}
\includegraphics[width=8.3cm]{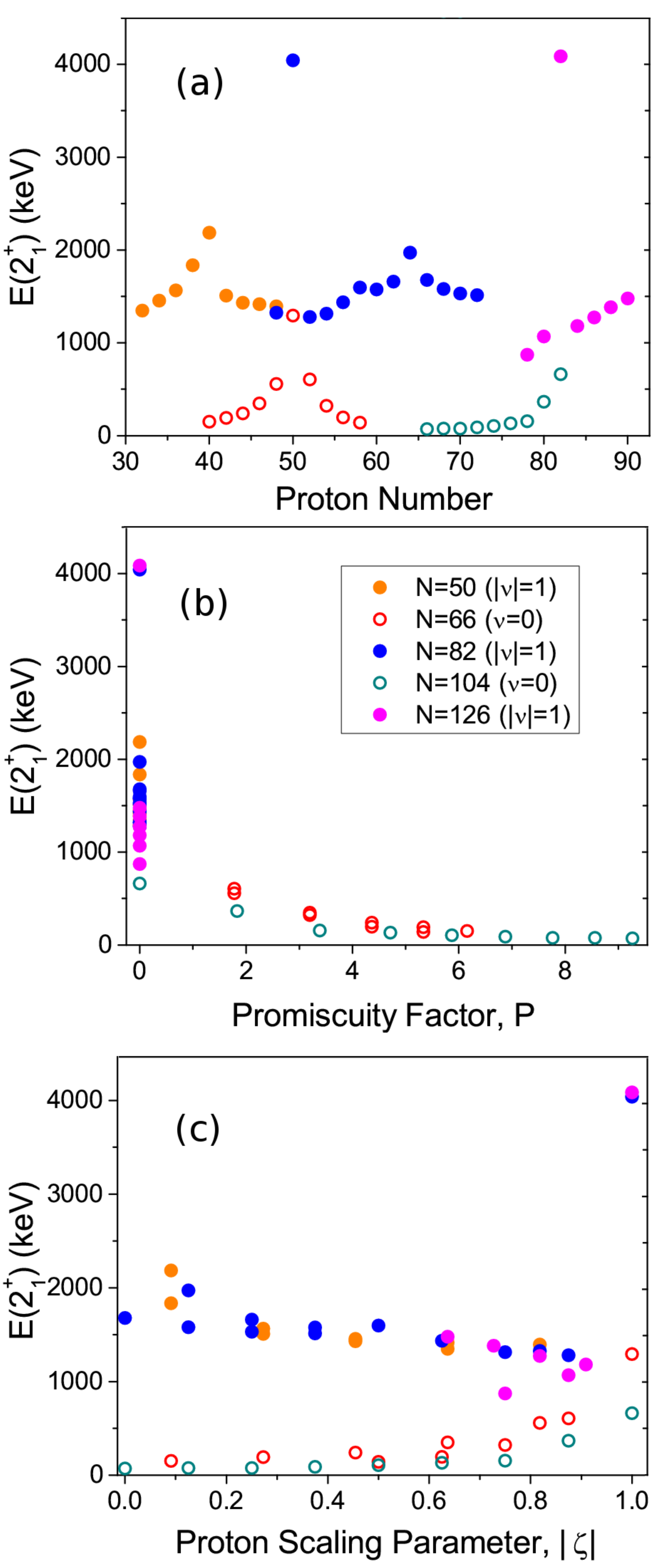}
\end{center}
\caption{ The energy of the $2_1^+$ state from Ref. \cite{E2} as a function of $Z$ in (a), $P$ in (b) and  $|\zeta|$ in (c). Filled circles are used for the chains in which the neutron number is magic and the hollow circles are used for 
nuclei that are in the exact middle of a given neutron shell. Please note that the value of  $P=0$ was given to doubly magic nuclei for which $P=0/0$.}
\label{fig:nu}

\end{figure}

The degree to which particle-hole symmetry exists can be demonstrated looking at experimentally
measured quantities like the energy of the $2_1^+$ state.
Nuclei with at least one shell closure can be seen to have larger $2_1^+$ energies. The $2_1^+$ energies are further enhanced if the nucleus is doubly magic. 
Fig. \ref{fig:nu} (a) shows energy of the $2_1^+$ state for five 
chains of isotones. 
Particle-hope symmetry will be invoked in the following section in order to simplify this figure.

\section{Particle-hole symmetry numbers}
The new formalism introduced in this manuscript can be defined in terms of the IBM-2 \cite{ibm2}, which distinguishes between protons and neutrons.
The boson number corresponds to the number of pairs of valence proton and neutron used, such that:
\begin{equation}
\label{Eqn:Npinu}
N_{\pi}=\frac{1}{2}(Z-Z_{min})\textrm{, and } N_{\nu}=\frac{1}{2}(N-N_{min}),
\end{equation}
which depend on magic numbers below or at the respective proton number ($Z$) and neutron number ($N$), denoted $Z_{min}$ and $N_{min}$. 
Similarly, valence boson holes can be defined as:
\begin{equation}
\label{Eqn:Npinubar}
\bar{N_{\pi}}=\frac{1}{2}(Z_{max}-Z)\textrm{, and } \bar{N_{\nu}}=\frac{1}{2}(N_{max}-N),
\end{equation}
which depend on the magic numbers at the end of a shell.

Valence particle-hole numbers, motivated by Eq. (\ref{rho}), are defined for neutrons and protons as:
\begin{equation}
\label{eqnu}
 \nu=\frac{N_{\nu}-\bar{N_{\nu}}}{N_{\nu}+\bar{N_{\nu}}}, 
\textrm{and }
 \zeta=\frac{N_{\pi}-\bar{N_{\pi}}}{N_{\pi}+\bar{N_{\pi}}}.
\end{equation}
The letters $\nu$ and $\zeta$ have been chosen to represent the valence particle-hole proton and neutron numbers in an attempt to follow the conventional use of N and Z.
Particle-hole symmetry can be invoked by using the absolute value or even powers of these numbers. 
Based on these definitions the numbers range between $-1$ and $1$ at the beginning and end of a shell and have a value at or near zero for mid-shell nuclides. 
 
 A comparison of the $E(2_1^+)$ as a function of the promiscuity factor and the absolute value of the new valence proton number are also provided in Fig. \ref{fig:nu} parts (b) and (c), respectively. 
The $P$-factor is shown here to be poorly correlated with the excitation energy for nuclei with one or two shell closures because $P=0$ in both cases. 
Comparable values of $|\zeta|$ correspond to similar $E(2_1^+)$ values for a given chain of isotones. This observation is evidence for particle-hole symmetry. Furthermore, 
there is also a separation of the nuclei with a closed shell of neutrons with $|\nu|$=1, compared to those which are mid-shell with $\nu$=0. 
The energies $2_1^+$ of nuclei with the same value of $|\zeta|$ for a given isotone are generally within $10\%$ of each other, and groups of nuclei with a constant value of $|\nu|$ and the same value of $|\zeta|$
are typically within $20\%$ of one another.
 
\section{Alternative representations}\label{AltRep}
F-spin provides another scheme within the IBM that can also be used to define $\nu$ and $\zeta$.
This also involves a classification of variables based on functions of the number of valence protons and neutrons, but now the total number of valence bosons:
\begin{equation} \label{Eqn:Nt}
N_T=N_{\pi }+N_{\nu }
\end{equation}
and the third projection of the F-spin:
\begin{equation}
\label{Eqn:F0}
 F_Z=\frac{1}{2}(N_{\pi}-N_{\nu}),
\end{equation}
are used which are known to commute with one another \cite{IBMReview2}.
The $F_Z$ number has been shown to allow for 
the prediction of various properties, from masses e.g. \cite{Davis91}, to spectroscopic factors e.g. \cite{Frank86}. 

There are also hole equivalents to $F_Z$ and $N_T$:
\begin{equation} \label{Eqn:NandFbar}
\bar{N_T}=\bar{N_{\pi}}+\bar{N_{\nu}} \textrm{, and } \bar{F_Z}=\frac{1}{2}(\bar{N_{\pi}}-\bar{N_{\nu}}),
\end{equation}
where the bar denotes holes as opposed to particles.
A combination of these terms have been used to describe energy levels in the ground state band of even-even nuclei \cite{dggrr}.
The valence particle-hole numbers are defined in terms of these F-spin related terms by: 
\begin{equation}
 \nu=\frac{N_T-\bar{N_T}-2F_Z+2\bar{F_Z}}{N_T+\bar{N_T}-2F_Z-2\bar{F_Z}} 
\textrm{, and }
 \zeta=\frac{N_T-\bar{N_T}+2F_Z-2\bar{F_Z}}{N_T+\bar{N_T}+2F_Z+2\bar{F_Z}}.
\end{equation}

The valence particle-hole numbers can also be generalized to include nuclei with odd numbers of nucleons.
This can be done by defining them in terms of proton and neutron numbers using the neighboring magic numbers, instead of using valence bosons:
\begin{equation} 
 \nu = \frac{2N-N_{max}-N_{min}}{N_{max}-N_{min}} 
\textrm{, and }
 \zeta = \frac{2Z-Z_{max}-Z_{min}}{Z_{max}-Z_{min}},
\end{equation}
where again the maximum and minimum values are defined by the nearest magic numbers.

Because the lack of experimental evidence of shell closures above $Z=82$ and $N=126$,
a theoretical calculation was used to predict the next major shell closure for protons and neutrons.
 The magic number $196$ for neutrons results 
from a Nilsson level calculation based on parameters from \cite{Bengtsson85}. Although, a 
substantial gap in the spherical levels was determined to exist at $184$, 
a gap roughly twice as large was found at $196$. This indicates that the $184$ gap, which has been suggested elsewhere in the literature e.g.
\cite{MHZ10}, should instead be treated as a sub-shell. A proton gap at $126$ also results from the 
calculation and is similar to the observed neutron shell closure at the same number. Therefore, initially, $Z_{min/max}=[2,8,20,28,50,82,126]$, and $N_{min/max}=[2,8,20,28,50,82,126,196]$ will be used as the proton and neutron magic numbers. 

\section{Test of particle-hole and valence proton-neutron symmetries in B(E2) values}\label{BE2test}

\begin{figure}
\begin{center}
\includegraphics[width=10cm]{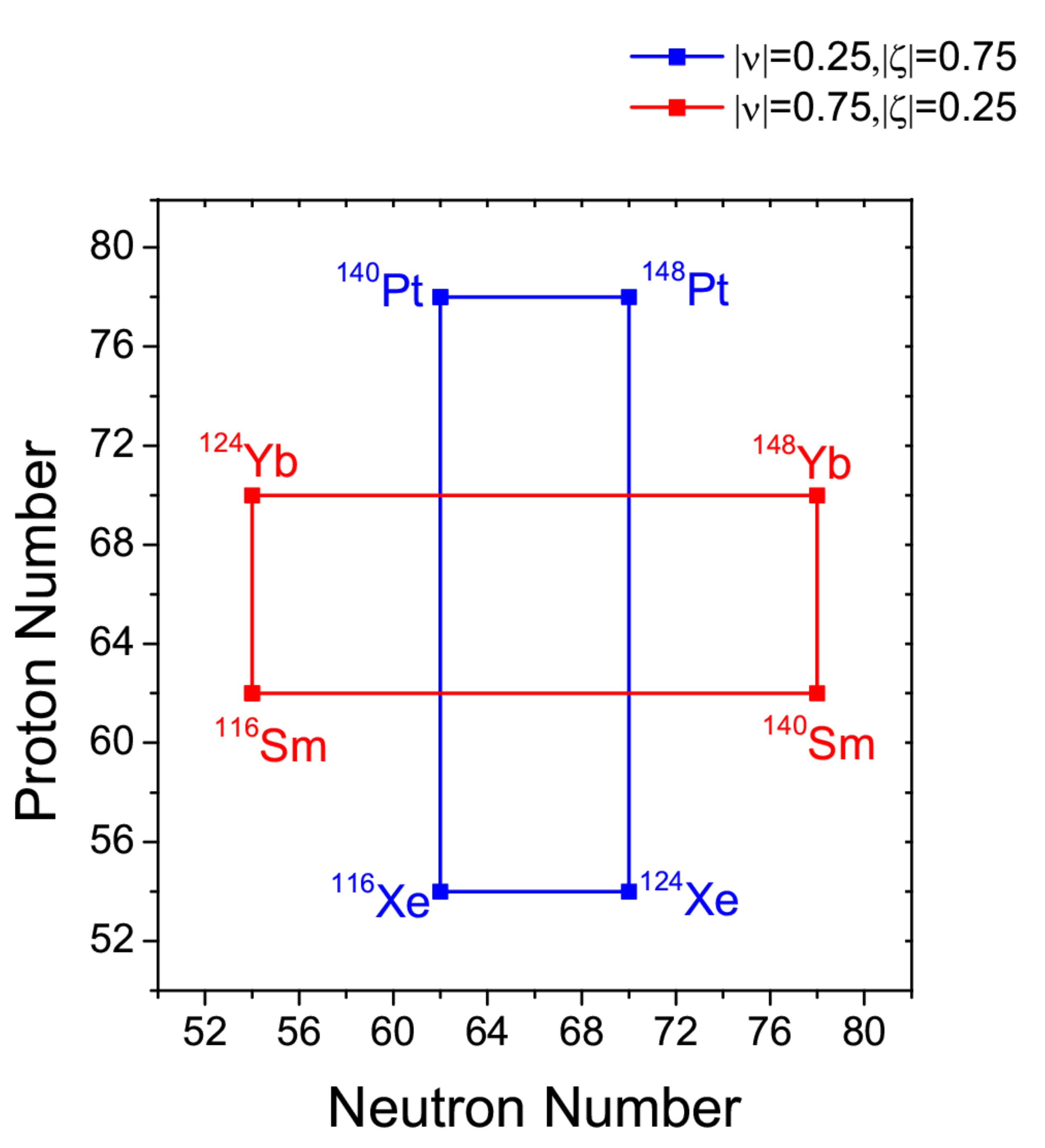}
\end{center}
\caption{The location of four-fold particle-hole symmetric multiplets denoted in red and blue. 
An octet comprised of the combination of these can be formed if valence proton-neutron symmetry is also invoked.}
\label{fig:Octet}
\end{figure}

The degree to which particle-hole symmetry exists can be demonstrated looking at experimentally
measured quantities such as the $B(E2)$ values in the ground state band of even-even nuclides.
Table \ref{BE2Table} contains $B(E2:0^+_1\rightarrow2^+_1)$ values from Ref. \cite{E2} for particle-hole symmetric nuclei with $50\leq N\leq82$ and $50\leq Z\leq82$.
Invoking particle-hole symmetry creates a four-fold multiplet of nuclei with the same $|\nu|$ and the same $|\zeta|$. 
A demonstration of two sample multiplets are given in Fig. \ref{fig:Octet}.

The coefficient of variation ($c_{B(E2)}$) can be used to compare $B(E2)$ values among the particle-hole symmetric nuclei. 
It is defined as:
\begin{equation}
 c_{B(E2)}=\sigma_{B(E2)}/ \bar{B}(E2),
\end{equation}
where $\bar{B}(E2)$ is the mean transition probability and $\sigma_{B(E2)}$ is the standard deviation about that mean.
The coefficient of variation metric demonstrates that on average this symmetry results in $c_{B(E2)}=13\%$ for the 11 groups of nuclei in this region. 

An additional proton-neutron mirror symmetry may exist as well, as has been investigated for many years, see e.g. \cite{Carlson}. In this work, mirror symmetry is extended and applied to valence groupings of nuclides and should not be confused with the more commonly discussed isospin symmetry.
Valence mirror symmetry can be tested by examining the $B(E2)$ values for two groups of four-fold multiplets where $|\nu|$ and $|\zeta|$ are swapped such as the two multiplets shown in Fig. \ref{fig:Octet}. 
The valence symmetry is tested using the $B(E2)$ values for two groups of four-fold multiplets with interchanged $|\nu|$ and $|\zeta|$, creating an octet of nuclei. The nineteen such groups, included in Table \ref{BE2Table2}, in the same mass region have 
on average $c_{B(E2)}=29\%$. It should be noted however that there are certainly cases shown where the symmetries are severely lacking, such as in the grouping of 
$^{126}$Ce and $^{138}$Gd which have $B(E2)$ values that vary by more than a factor of five from each other \cite{E2}.
Although neither of these symmetries are exact, the approximate nature of both symmetries provides an opportunity to model global shell features of nuclear binding energies and energy ratios within the ground state band.

\section{Results and discussion}
 \label{Sec.fittype}

The physical understanding for the success of these new numbers has to do with the shape of the nuclides and the corresponding collective behavior. 
In general, nuclides near one or more shell closure are spherical and have vibrational structure, this results in a low $B(E2)$ and $R_{4/2}$, 
where as mid-shell nuclides are well deformed and their rotational behavior generates higher $B(E2)$ and $R_{4/2}$ values.
Generally speaking high values of either $|\nu|$ or $|\zeta|$ will correspond to vibrational nuclei and low values of both correspond to rotational nuclei. 
 
The approximate valence proton-neutron symmetry as seen in transition rates can be used to justify global fits consisting of the same leading powers of $\nu$ and $\zeta$.
Further, combinations of even powers of these numbers can be used to invoke particle-hole symmetry. One combination that satisfies these two criteria is of the form $(\nu^4+c\nu^2\zeta^2+\zeta^4)$, where $c$ is a constant.
Combinations of this form are used to empirically model shell effects for two experimental observables across the chart of the nuclides.

\subsection{$4^+_1/2^+_1$ ratio fit} \label{4to2}

\begin{figure}
\begin{center}
\includegraphics[width=9.7cm]{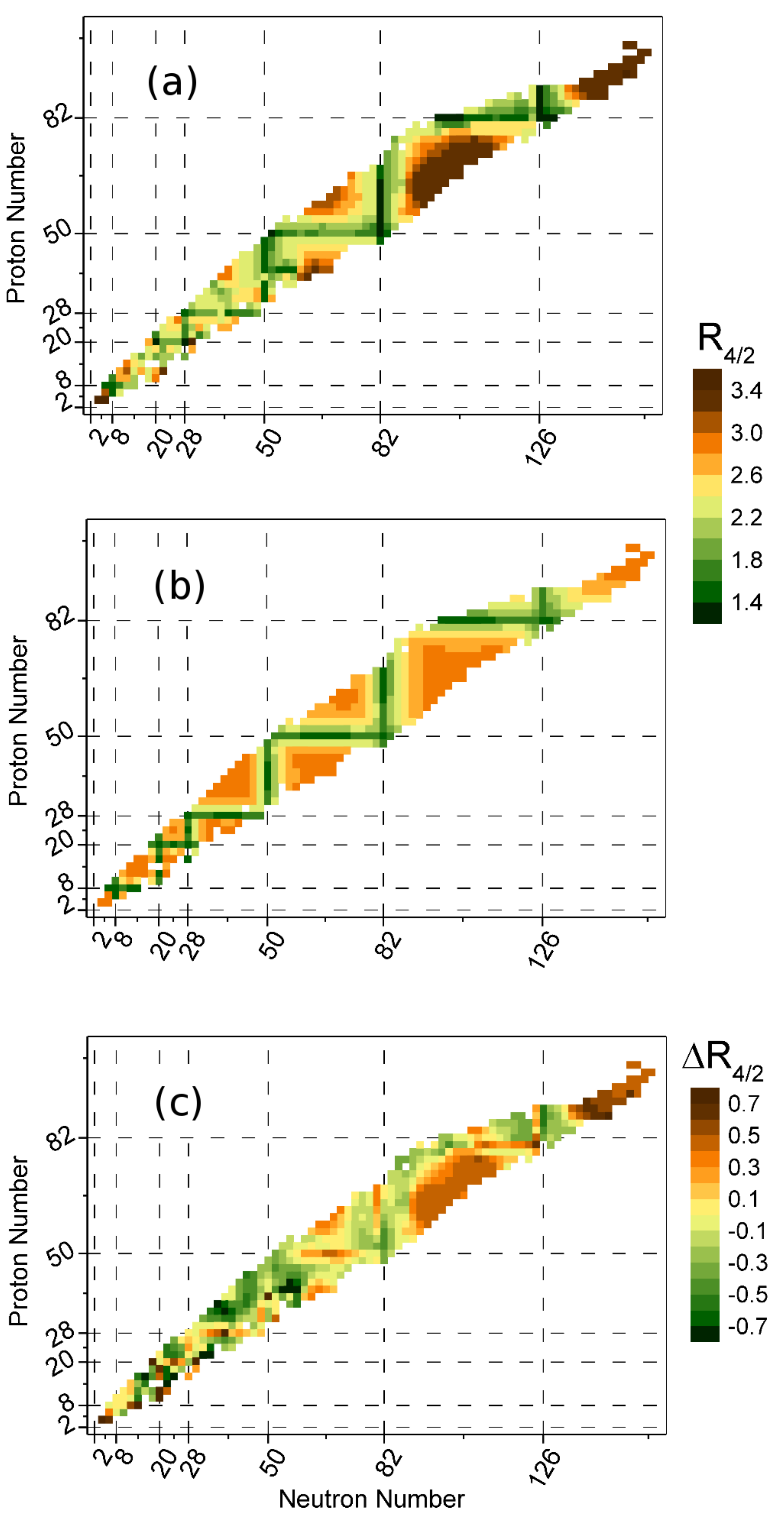}
\end{center}
\caption{ The experimental $R_{4/2}$ values from \cite{E2} in (a) and the two parameter fit to the $R_{4/2}$ using Eq. (\ref{Eq:R4/2}) in (b), as well as the difference between experiment and theory in (c). Please note that (a) and (b) share the same range scale for coloring and (c) uses a different smaller range scale.}
\label{fig:4to2}
\end{figure} 

Fig. \ref{fig:4to2} (a) shows the $R_{4/2}$ values which come to a maximum
value of approximately 3.33 at mid-shell regions and go below a value of 2. 
The mean value of the $R_{4/2}$ is 2.40, for measurements of 557 even-even nuclei in Ref. \cite{E2} and the standard
deviation about that mean value is $\sigma=0.58$.

Various linear combinations of even powers of $\nu$ and $\zeta$ were systematically tested and the combination $\nu^4-\nu^2\zeta^2+\zeta^4$ generated the optimal standard deviation for a fit of $R_{4/2}$.
This is justified based on empirical observations of phenomena in the closed shell regions. 
A function comprised of the sum of  $\nu^4$ and $\zeta^4$ the total
is twice as large for a doubly magic nucleus
as it is for a singly magic nucleus.  
Subtracting a cross term of the form, $\nu^2\zeta^2$ causes the value at
with one shell closure to be
equal to the value with two shell closures. In this way, the combination works similarly to $P$ and correlates collectivity with $R_{4/2}$.

The corresponding best fit is therefore of the form: 
\begin{equation}\label{Eq:R4/2}
R_{4/2,fit}=R_{shell}(\nu^4-\nu^2\zeta^2+\zeta^4)+R_{max},
\end{equation}
with $R_{shell}=$-1.46, and $R_{max}=$2.88. The resulting values are included in Fig. \ref{fig:4to2}. 
This fit has reduced the standard deviation to
$\sigma=0.37$.

A shape phase transition, as indicated by $R_{4/2}$ takes place at $\nu^4-\nu^2\zeta^2+\zeta^4 \approx 0.33$. Values of $\nu^4-\nu^2\zeta^2+\zeta^4$ less than this and ranging down to zero are generally rotational and values greater than this, up to a value of one, are generally vibrational.

The difference $\Delta R_{4/2}=R_{4/2,exp.}-R_{4/2,fit}$ is shown in Fig. \ref{fig:4to2} (c). This difference can be used as an indicator of sub-shell structure.
The $R_{4/2}$ metric provides insight on cases such as the N=32 sub-shell closure that corresponds to a low $2^+_1$ excitation seen experimentally for titanium and chromium \cite{Hagen12}. This 
sub-shell closure corresponds to a low $4^+_1$ state in titanium, but not in chromium as can be inferred from Fig. \ref{fig:4to2} (c) where $^{54}$Ti is deviates from the trend but $^{56}$Cr does not. 
Furthermore, $\Delta R_{4/2}$ can be used to demonstrate that the $R_{4/2}$ for nuclei including $^{30}$Ne, $^{32}$Mg, $^{32}$S, $^{38,50}$Ca, $^{54}$Ti, $^{66}$Ni, $^{72}$Kr, 
$^{88}$Sr, and $^{96,98}$Zr deviate substantially from the globally fit trends. In some cases this may suggest that further measurements are needed or this may provide motivation for future theoretical investigations of these nuclei.

\subsection{Semi-empirical binding energy fit}  \label{BE}
\begin{figure}
\begin{center}
\includegraphics[width=9.7cm]{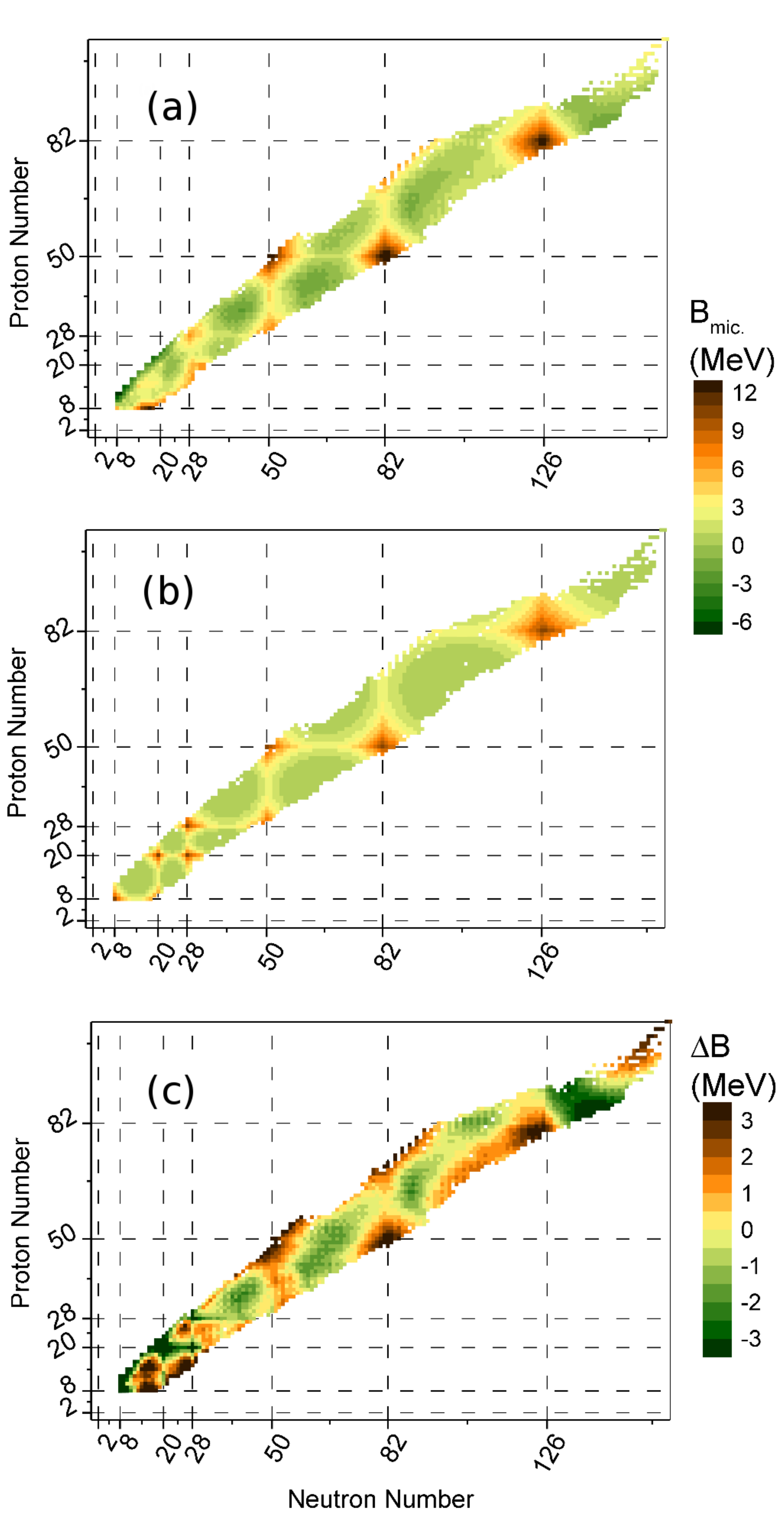}
\end{center}
\caption{ (a) The experimental
\textquotedblleft microscopic$\textquotedblright$ binding energies
determined using Eq. (\ref{Eq:micExp}) based on data from \cite{Au12} with the electron binding energy contribution has been removed. (b) The theoretical shell contribution fit using Eq.
(\ref{Eq:micFit}). (c) The difference between the experimental
and theoretical values. Please note that (a) and (b) share the same range and that (c) uses a smaller range scale. }
\label{fig:BE}
\end{figure} 

Binding energies are inherently anti-symmetric with regard to isospin and particle-hole symmetry as a result of the 
mutual charge repulsion of protons represented by the Coulomb term. In order to isolate the particle-hole symmetric
term, a semi-empirical binding energy fit will be removed from experimental binding energies, which contains Coulomb and isospin contributions.
Such semi-empirical binding energy formulas have existed for many years, see e.g. \cite{Wiez35}, and they often treat the nucleus as a charged droplet of nuclear matter.
The terms in the semi-empirical formula from Myers and Swiatecki \cite{MS66} is used as a
starting point for binding energy fits.  
The five term fit used to model \textquotedblleft macroscopic$\textquotedblright$ binding energy contributions is:
\begin{equation}\label{Eq:macFit} 
B_{mac.,th.} = ( a_v A+a_s A^{2/3}) ( 1+\kappa T_Z(T_Z+1)A^{-2}) + (a_c Z(Z-1)+ \Delta) A^{-1/3},
\end{equation} 
where $A=N+Z$ is the total number of nucleons and $T_Z=(N-Z)/2$ is the isospin projection. Here, the dependence of the Coulomb, pairing and symmetry energy terms have all been modified from the original Myers and Swiatecki expression. 

 The pairing contribution used is $\Delta=+a_p$ if the 
nucleus is even-even, $\Delta=-a_p$ if odd-odd, and is $\Delta=0$ otherwise. The use of a pairing term proportional to $A^{-1/3}$ suggested by Ref. \cite{MS66} provides a minor improvement to the overall standard deviation on the order of
10-20 keV compared to the more commonly used $A^{-1/2}$ dependence. This finding is in contradiction to Ref. \cite{MT08} which states that no distinction was found when comparing the pairing term's dependence on A. The $Z(Z-1)$ expansion results
from the semi-classical treatment of the protons accounting for the interaction of protons with other protons. The expansion of the form $T_Z(T_Z+X)$ is consistent with experimental observations for $N\approx Z$ nuclei \cite{Ja02},
with deviations from $X=1$ likely resulting from changes in level density near the Fermi surface as discussed in Ref. \cite{Bentley13}. A sixth term discussed below accounts for shell effects in an approximate way. 

To allow for better understanding of the shell term, the macroscopic binding energy fit can be removed from the experimental binding energies leaving
just the \textquotedblleft experimental microscopic$\textquotedblright$ contribution, such that:
\begin{equation} \label{Eq:micExp}
B_{mic.,ex.}=B_{ex.}-B_{mac.,th.},
\end{equation}
which is comparable to the use of semi-empirical microscopic masses discussed in Ref. \cite{Hau88}.
The periodicity of the experimental microscopic binding energy contribution has long been known to be a consequence of shell effects, see e.g. \cite{MS66}, and is shown in Fig. \ref{fig:BE} (a).

Various linear combinations of even powers of $\nu$ and $\zeta$ were also systematically tested and the combination $\nu^4+2\nu^2\zeta^2+\zeta^4$ was determined to generate the optimal fit of the \textquotedblleft microscopic$\textquotedblright$ binding energy component.
An empirical justification for the cross-term coefficient can be seen in Fig. \ref{fig:BE} (a) which contains values of approximately 3 MeV when one shell is closed and approximately 12 MeV when both are closed.
For this reason the \textquotedblleft microscopic$\textquotedblright$ binding energy component has a value that is approximately four times larger when both shells are closed than if just one is closed, as can be modeled by:
\begin{equation} \label{Eq:micFit}
B_{mic.,th.}=a_{shell}(\nu^2+\zeta^2)^2,
\end{equation}
where $a_{shell}$ is the sixth adjustable coefficient. Physically, this term represents deformation effects including the enhanced pairing correlations calculated micro-macro models resulting from large gaps in the single particle levels at a closed shell.

The best fits discussed below were determined by performing a $\chi^2$ minimization using OriginPro 9.0 \cite{Origin9} followed by a standard deviation ($\sigma$) minimization. 
The six terms generate a fit with a value of $\sigma=$1.71 MeV for 2353 nuclei in the range $N>8$ and $Z>8$ found in the 2012 Atomic Mass Evaluation \cite{Au12} 
with the electron binding energy contribution removed using Eqn. (A4) from Ref. \cite{Lunney03}. The values of $B_{mic.,ex.}$ and $B_{mic.,th.}$ resulting from using these coefficients are demonstrated in Fig. \ref{fig:BE} (b). The difference $\Delta B=B_{exp.}-B_{fit}$, where $B_{fit}=B_{mac.,th.}+B_{mic.,th.}$, is shown in Fig. \ref{fig:BE} (c).

Following the work discussed in Ref. \cite{VanIsackerGlobal} additional improvements were made to the binding energy fit by changing the conventional $N,Z=20$ shell closures for $N,Z=14$ as has been suggested by Dieperink and Van Isacker \cite{DVI09}
and  $Z=114$  in place of $Z=126$ as suggested by theoretical calculations as discussed in Ref. \cite{Z114}. These changes in magic numbers require only a slight 
modification of some of the fit coefficients to create the new best fit with a standard deviation of $\sigma=$1.55 MeV. The coefficients corresponding to this fit are $a_v=$15.91 MeV,  $a_s=$-18.65 MeV, $\kappa=$-7.26, $a_c=$-0.7197 MeV, $a_p=$2.91 MeV and $a_{shell}=$5.00 MeV. 

For some perspective, one of the alternate shell corrections tested, involved a shell term of the form $a_{shell}(\nu^2+\zeta^2)$ and resulted in a standard deviation of $\sigma=$1.79 MeV. This finding is consistent with a general observation that the results were not particularly sensitive to the powers to which $\nu$ and $\zeta$ are raised.

To compare the importance of shell and pairing terms on the binding energy, fits were also performed
in which either the shell or pairing coefficients were set equal to zero and the remaining five terms were adjusted.
For $a_{shell}=0$, the best fit resulted in a standard deviation of $\sigma=$2.65 MeV as opposed to 
the case with $a_p=0$, where the best fit resulted in $\sigma=$1.88 MeV. These results indicate that the pairing gap, which is commonly used in semi-empirical formulae, is less critical for modeling binding energies than shell effects, which is often rarely accounted for.

\section{Conclusions} \label{outlook}

The $\nu$ and $\zeta$ numbers, defined in Eq. (\ref{eqnu}), have been introduced to allow for a relatively simple modeling of nuclear shell effects.
All nuclei to fit between the range of -1 and 1 in terms of the new numbers. 
The numbers can be used to predict observables based on the behavior of nuclei at comparable locations in the same shell or in adjacent shells. 
$B(E2)$ values indicate that particle-hole and valence proton-neutron symmetries are approximate ($10-30\%$) in the range $50 \leq N,Z \leq 82$. 

Particle-hole and valence proton-neutron symmetries can continue to be explored using these numbers.
As opposed to focusing on multiple parameter fits, even powers of these numbers have been 
combined to model global features of binding energies and $R_{4/2}$ over the chart of the nuclides.
After fixing the respective $\nu^2\zeta^2$ cross terms, only one fit parameter is required.

Overall, the $P$ factor is the preferable variable to use with $R_{4/2}$, as opposed to Eq. (\ref{Eq:R4/2}), because of its proven success and relative ease of use. It should be noted, however, that 
$P$ cannot be used to provide a sufficient description 
of the constitute $E(2_1^+)$ and $E(4_1^+)$, $B(E2)$ values or binding energies because it cannot distinguish between singly and doubly magic nuclei. 
Fits using $F_Z$ or other related quantities which do make the appropriate distinction, 
often require multiple fit coefficients, see e.g. \cite{Davis91} or \cite{dggrr}.

For binding energies, the inclusion of a single, relatively simple, shell term has reduced the standard deviation of semi-empirical fits by more than 1 MeV.  The fit using Eqs. (\ref{Eq:macFit}) and (\ref{Eq:micFit}) and can produce binding energies of 2353 nuclei with a standard deviation of $\sigma=$1.55 MeV.

Future studies of various nuclear properties including: energies of excited states, nuclear charge radii, nuclear magnetic moments and drip-lines can be performed using this formalism.

\ack{The author would like to thank Ani Aprahamian for supporting this project and Ana Georgieva for pointing out the relations between the new numbers and boson operators.}

\begin{table}
\begin{center}
\caption{Experimental $B(E2)$ values, in units of $e^2b^2$, from Ref. \cite{E2} for nuclei with the same $|\nu|$ and $|\zeta|$ in the region $50 \leq N,Z \leq 82$. \label{BE2Table}}
\begin{tabular}{cc|cccccc|c} %\toprule
 $|\nu|$ &  $|\zeta|$ & $^{ A }$X & B(E2) &Error & $^{ A }$X & B(E2) &Error &  $c_{B(E2)}$ \\
 \hline
0.125&0.750&$^{118}$Xe&1.4&(0.1)&$^{122}$Xe&1.39&(0.08)&0.005\\
0.125&1.000&$^{114}$Sn&0.232&(0.008)&$^{118}$Sn&0.183&(0.009)&0.167\\
0.250&0.750&$^{116}$Xe&1.21&(0.06)&$^{124}$Xe&1.12&(0.15)&0.055\\
0.250&0.875&$^{114}$Te&0.611&(0.054)&$^{122}$Te&0.6685&(0.0046)&0.064\\
0.250&1.000&$^{112}$Sn&0.242&(0.008)&$^{120}$Sn&0.191&(0.01)&0.167\\
0.375&0.750&$^{114}$Xe&1.033&(0.0666)&$^{126}$Xe&1.02&(0.14)&0.009\\
0.375&1.000&$^{110}$Sn&0.22&(0.022)&$^{122}$Sn&0.164&(0.01)&0.206\\
0.500&1.000&$^{108}$Sn&0.222&(0.019)&$^{124}$Sn&0.14&(0.01)&0.320\\
0.625&0.875&$^{108}$Te&0.39&(0.045)&$^{128}$Te&0.3755&(0.0033)&0.027\\
0.625&1.000&$^{106}$Sn&0.195&(0.039)&$^{126}$Sn&0.14&(0.02)&0.232\\
0.750&1.000&$^{104}$Sn&0.1&(0.04)&$^{128}$Sn&0.08&(0.005)&0.157\\

\hline                               
\end{tabular}
\end{center}
\end{table}

\begin{table}
\begin{center}
\caption{Experimental $B(E2)$ values, in units of $e^2b^2$, from Ref. \cite{E2} for nuclei with $|\nu|\rightarrow|\zeta|$ and $|\zeta|\rightarrow|\nu|$ in the region $50 \leq N,Z \leq 82$. \label{BE2Table2}}
\begin{tabular}{cc|cccccc|c} %\toprule
 $|\nu|$ &  $|\zeta|$ & $^{ A }$X & B(E2) & Error &$^{ A }$X & B(E2) &  Error &$c_{B(E2)}$  \\
 \hline
0.125&0.500&$^{126}$Ce&4.1&(0.8)&&&&1.312\\
0.500&0.125&$^{138}$Gd&0.1542&(0.0085)&&&&\\
\hline
0.250&0.375&$^{130}$Nd&4.1&(1.8)&&&&0.017\\
0.375&0.250&$^{134}$Sm&4.2&(0.6)&&&&\\
\hline
0.250&0.500&$^{128}$Ce&2.4&(0.25)&&&&0.091\\
0.500&0.250&$^{136}$Sm&2.73&(0.27)&&&&\\
\hline
0.250&0.625&$^{126}$Ba&1.69&(0.05)&&&&0.223\\
0.625&0.250&$^{138}$Sm&1.23&(0.17)&&&&\\
\hline
0.250&1.000&$^{112}$Sn&0.242&(0.008)&$^{120}$Sn&0.191&(0.01)&0.156\\
1.000&0.250&$^{144}$Sm&0.261&(0.011)&&&&\\
\hline
0.375&0.500&$^{130}$Ce&2&(0.17)&&&&0.077\\
0.500&0.375&$^{134}$Nd&2.23&(0.29)&&&&\\
\hline
0.375&0.625&$^{128}$Ba&1.375&(0.12)&&&&0.133\\
0.625&0.375&$^{136}$Nd&1.6615&(0.23)&&&&\\
\hline
0.375&0.875&$^{124}$Te&0.561&(0.024)&&&&0.176\\
0.875&0.375&$^{140}$Nd&0.72&(0.05)&&&&\\
\hline
0.375&1.000&$^{110}$Sn&0.22&(0.022)&$^{122}$Sn&0.164&(0.01)&0.355\\
1.000&0.375&$^{142}$Nd&0.33&(0.09)&&&&\\
\hline
0.500&0.625&$^{130}$Ba&1.1&(0.015)&&&&0.013\\
0.625&0.500&$^{134}$Ce&1.08&(0.09)&&&&\\
\hline
0.625&0.750&$^{130}$Xe&0.585&(0.011)&&&&0.080\\
0.750&0.625&$^{134}$Ba&0.655&(0.006)&&&&\\
\hline
0.500&0.750&$^{128}$Xe&0.9095&(0.2217)&&&&0.082\\
0.750&0.500&$^{136}$Ce&0.81&(0.09)&&&&\\
\hline
0.500&0.875&$^{126}$Te&0.457&(0.014)&&&&0.005\\
0.875&0.500&$^{138}$Ce&0.46&(0.05)&&&&\\
\hline
0.500&1.000&$^{108}$Sn&0.222&(0.019)&$^{124}$Sn&0.14&(0.01)&0.493\\
1.000&0.500&$^{140}$Ce&0.38&(0.11)&&&&\\
\hline
0.625&0.875&$^{108}$Te&0.39&(0.045)&$^{128}$Te&0.3755&(0.0033)&0.111\\
0.875&0.625&$^{136}$Ba&0.46&(0.04)&&&&\\
\hline
0.625&1.000&$^{106}$Sn&0.195&(0.039)&$^{126}$Sn&0.14&(0.02)&0.282\\
1.000&0.625&$^{138}$Ba&0.25&(0.1)&&&&\\
\hline
0.750&0.875&$^{130}$Te&0.295&(0.007)&&&&0.062\\
0.875&0.750&$^{134}$Xe&0.322&(0.044)&&&&\\
\hline
0.750&1.000&$^{104}$Sn&0.1&(0.04)&$^{128}$Sn&0.08&(0.005)&0.868\\
1.000&0.750&$^{136}$Xe&0.36&(0.06)&&&&\\
\hline
0.875&1.000&$^{130}$Sn&0.023&(0.005)&&&&0.939\\
1.000&0.875&$^{134}$Te&0.114&(0.013)&&&&\\

\hline

\end{tabular}
\end{center}
\end{table}

\end{document}